\documentclass[aps,prl,reprint,groupedaddress]{revtex4-1}
\usepackage{amssymb,amsfonts,amsmath}
\usepackage[dvips]{graphicx,color,psfrag}

\begin{document}


\title{Disadvantages of Preferential Dispersals in Fluctuating Environments}


\author{Satoru Morita}
\email[]{morita@sys.eng.shizuoka.ac.jp}
\affiliation{Department of Mathematical and Systems Engineering, Shizuoka University, Hamamatsu, 432-8561, Japan}
\author{Jin Yoshimura}
\email[]{jin@sys.eng.shizuoka.ac.jp}
\affiliation{Department of Mathematical and Systems Engineering, Shizuoka University, Hamamatsu, 432-8561, Japan}
\affiliation{Marine Biosystems Research Center, Chiba University, Uchiura, Kamogawa, Chiba 299-5502, Japan}
\affiliation{Department of Environmental and Forest Biology, State University of New York College of Environmental Science and Forestry, Syracuse, NY 13210 USA}


\date{\today}

\begin{abstract}
It has not been known 
whether preferential dispersal is adaptive in fluctuating environments
We investigate the effect of preferential 
and random dispersals in bet-hedging systems 
by using a discrete stochastic metapopulation model,
where each site fluctuates between good and bad environments 
with temporal correlation. 
To explore the optimal migration pattern, 
an analytical estimation of the total growth 
is derived by mean field approximation.
We found that the preference for fertile sites is disadvantageous
when transportation among sites has a cost or
the sensitivity of preference is high. 
\end{abstract}


\maketitle



Bet-hedging is a risk spreading strategy 
that increases a population's fitness 
in the face of fluctuating environments \cite{seger,r1,r2,iwasa,sasaki,childs,r3}.
For example, consider offspring allocation into two sites, 
either of which is so poor that 
the population cannot survive without dispersal \cite{r4,r8}. 
The random dispersal between two sites can lead to 
population persistence.
Also in finance, diversified investment is a key point 
for risk mitigation \cite{r10}.
To increase expected returns, the wealth should be invested 
into a wide variety of assets.
In these bet-hedging models, random dispersal is assumed.

However, in the case that the environment
of each site fluctuates with temporal correlation,
it is not clear whether
the random dispersal is the best strategy as bet-hedging.
There is a possibility that the preferential dispersal improves the growth 
of a population.
If the environment has a temporal correlation,
a large population implies that  environment tends to 
remain good subsequently.
Therefore, for a individual changing the habitat,
it is advantageous to choose populous sites 
than sparsely populated ones. 
However, what is good for individual is not necessarily 
good for a whole.

In this paper, we study a stochastic metapopulation
which is given as a discrete-time stochastic matrix model \cite{r7,morita}.
In these studies,
to evaluate the population's fitness for this class model,
the total growth was calculated approximately. 
Assuming random dispersal, the preferential dispersal has not been taken into consideration.
Here, we introduce a simple model with preferential dispersal
and present the mean-field theory for the model.
The purpose of this paper is to compare the total population growths between preferential and random dispersals

Let us consider population that inhabits in $n$ discrete sites. 
Let $x_i (t)$ be the number of individuals in site $i \ (1\leq i\leq n)$
at time $t$.
Thus, the sum of populations in all sites is 
\begin{equation}
S(t)=\sum_{i=1}^n  x_i(t) .
\label{e01}
\end{equation}
The initial condition is set as $x_i(0)=1$, i.e., $S(0)=n$.
In each site, the population reproduces with random growth 
$m_i (t)$, which are stochastic variables with 
finite variance. 
The probability distributions of $m_i (t)$ are assumed be the 
same for all sites. 
The total growth is given as
\begin{equation}
\bar{m}(t)=\frac{1}{S(t)}\sum_{i=1}^n  m_i(t)x_i(t).
\label{e02}
\end{equation}
Thus, 
\begin{equation}
S(t+1)=\bar{m}(t)S(t).
\label{e05}
\end{equation}

We assume that all sites are connected to one another, 
and a proportion $q$ of the population tries to migrate at each time step.
The remaining population stays at the same site.
Thus, the population dynamics is written as 
\begin{equation}
 x_i(t+1)=(1-q)m_i(t)x_i(t)+q \bar{m}(t)S(t) F(\{x_i(t)\}),
\label{e03}
\end{equation}
where $q$ represents the migration rate,
$q \bar{m}(t)S(t)$ is the number of population that migrates,
and $F(\cdot)$ represents the preference of migration.
We assume that
a fraction $r$ of the migrating population 
tends to be attracted to fertile sites
and the remain moves at random. 
Then, we have
\begin{equation}
F(\{x_i(t)\})=r\frac{m_i x_i(t)}{\bar{m}(t)S(t)}+(1-r)\frac{1}{n}.
\label{e04}
\end{equation}
The first term of the right hand side of this equation means
that the fraction $r$ of the migrating population
selects sites with the probability linearly proportional to the population. 
In summary, each individual 
stays the same site with the probability $(1-q)$,
migrates preferentially with the probability $qr$,
and migrates randomly with the probability $q(1-r)$.
Summing \eqref{e03} over all sites, we obtain \eqref{e05}, again.
Let us denote the relative population size by
\begin{equation}
y_i(t)=\frac{x_i(t)}{S(t)}.
\label{e06}
\end{equation}
Dividing both sides of \eqref{e03} by both sides of \eqref{e05} and
using \eqref{e04},
the dynamics of the relative population size $y_i(t)$ is obtained
\begin{equation}
y_i(t+1)=\frac{m_i(t)}{\bar{m}(t)}(1-q+qr)y_i(t)+\frac{1}{n}q(1-r).
\label{e07}
\end{equation}
Since eq.~\eqref{e02} is rewritten as
\begin{equation}
\bar{m}(t)=\sum_{i=1}^n  m_i(t)y_i(t),
\end{equation}
the stochastic process \eqref{e07} describes 
the evolution of $y_i(t)$ completely.
It is obvious that the condition
$
\sum_{i=1}^{n}y_i(t)=1
$
is conserved for the dynamics \eqref{e07}.

For simplicity,
the local growth $m_i(t)$ is assumed to take one of two values, 
$m_-$ with probability $p$ and $m_+$ with probability $1-p$.
Here, we set $m_+>m_-$.
The stochastic fluctuation of the local growth comes from 
environmental fluctuations. 
The cases of $m_+$ and $m_-$ are called good and bad environments,
respectively.
Here, we take into account temporal correlation of $m_i(t)$.
We introduce the autocorrelation function 
\begin{equation}
R(\tau)=\langle m_i(t)m_i(t+\tau)\rangle-
\langle m_i(t)\rangle \langle m_i(t+\tau)\rangle
\label{e10}
\end{equation}
that decays exponentially 
\begin{equation}
R(\tau)\propto \exp(-\tau/\tau_c),
\label{e11}
\end{equation}
where $\tau_c$ is an autocorrelation time.
To obtain a stochastic time series satisfying the property \eqref{e11}, 
we use Markov chain with
transition probability matrix  
\begin{equation}
P_{env}=
\left(\begin{array}{ll}
1-pu & pu\\
(1-p)u & 1-(1-p)u
\end{array}
\right) ,
\label{e12}
\end{equation}
where 
\begin{equation}
u=1-\exp(-1/\tau_c).
\end{equation}
For example, 
the first row and second column element of the matrix $P_{env}$ represents 
the probability that a site changes from good to bad.
Here, we neglect spatial correlation of the local growth $m_i (t)$
for simplicity.

Before analyzing the model, we remark that 
since the stochastic process \eqref{e07}
depends on $q$ and $r$ only through the form $q(1-r)$,
the solution of \eqref{e07} is identical for the same vales of $q(1-r)$,
Thus, it is convenient to introduce a new combined parameter 
\begin{equation}
\phi=q(1-r).
\end{equation}

Now, we develop the mean-field approximation for the model.
The ratios of population in the two 
environments are denoted as $\rho_+(t)$ and $\rho_-(t)$, 
respectively.
Here, $\rho_+(t) +\rho_-(t)=1$ and 
the total growth is given as 
\begin{equation}
\bar{m}(t)=m_+\rho_+(t)+m_-\rho_-(t).
\end{equation}
%
%
The evolution of $(\rho_+(t),\rho_-(t))$
is described by 
\begin{equation}
(\rho_+(t+1),\rho_-(t+1))=
\left(
\frac{m_+}{\bar{m}(t)}\rho_+(t),\frac{m_-}{\bar{m}(t)}\rho_-(t)
\right)
P_{all}.
\label{e21}
\end{equation}
Here, the term $P_{all}$ is defined 
\begin{equation}
P_{all}=(1-q)P_{env}+q(1-r)P_{rand}+qrP_{pref}P_{env},
\label{e18}
\end{equation}
where
\begin{equation}
P_{rand}=\left(\begin{array}{ll}
1-p & p\\
1-p & p
\end{array}
\right)
\end{equation}
and 
\begin{equation}
P_{pref}=
\frac{1}{\bar{m}(t)}\left(\begin{array}{ll}
m_+ \rho_+(t) & m_- \rho_-(t)\\
m_+ \rho_+(t) & m_- \rho_-(t)
\end{array}
\right).
\end{equation}
The first term of the right hand side of \eqref{e18}
stands for the case 
that the individual continues to stay at the same site
and the environment changes.
The second term of \eqref{e18} stands for the case that
the individual moves randomly.
The first (second) column of matrix $P_{rand}$ 
is the probabilities that 
the environments in the end of the migration
is good (bad).
The final term of \eqref{e18} stands for 
the case of  preferential dispersal.
The first (second) column of matrix $P_{pref}$ 
is the probabilities that 
the environments in the end of the migration
is good (bad).

This approximation 
is true in the limit of $n\to \infty$.
A stationary solution $\rho_+=\rho_+(t+1)=\rho_+(t)$,
is obtained analytically by working out \eqref{e21}.
Because $P_{all}$ and $\bar{m}(t)$ depend on $\rho_+(t)$ and $\rho_-(t)$,
we need to solve a quadratic equation. 
By solving \eqref{e21}, the expected value of the total growth 
is expressed as 
\begin{equation}
\bar{m}=m_+ \rho_+ + m_- \rho_- ,
\label{e22}
\end{equation}
where $\rho_+$ and $\rho_-$ are its solutions.
In the absence of temporal correlation ($\tau_c=0$),
\eqref{e22} is rewritten as $\bar{m}=m_+ (1-p) + m_- p$.
In this case, the total growth is independent of $q$ and $r$.
On the other hand, if $\tau_c\neq 0$,
the total growth takes an intricate form.

\begin{figure}
\begin{center}
\psfrag{q(1-r)}{{\small $\phi$}}
\psfrag{total growth rate}{{\small $\log\bar{m}$}}
\psfrag{tau = 2}{$\tau_c=2$}
\psfrag{tau = 1}{$\tau_c=1$}
\psfrag{tau = 0.5}{$\tau_c=0.5$}
\psfrag{tau = 0}{$\tau_c=0$}
\psfrag{q}{{\small $q$}}
\psfrag{r}{{\small $r$}}
\includegraphics[width=0.5\textwidth]{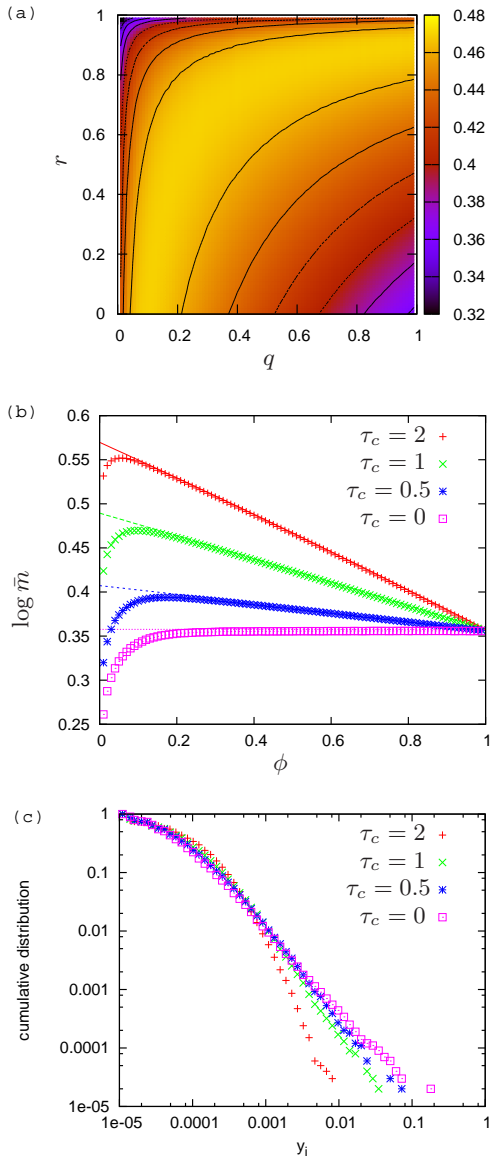}\\%
\end{center}
\caption{ 
(a) The dependence of 
the population's fitness of 
$q$ and $r$.
The average of total growth rate $\log \bar{m}$ is 
calculated by numerical simulation for 10,000 time steps
over a transient of 10,000 time steps, when $\tau_c=1$.
The plots are given by the average over 100 stochastic realizations.
The other parameters are set as $m_+=2$, $m_-=0.1$, $p=0.3$, $n=100$.
(b) The average of total growth rate of as a function of $\phi=q(1-r)$
for three values of autocorrelation time $\tau_c=0, 0.5, 1, 2$.
The crosses represent the numerical result for $r=0$.
The curves give the prediction by the mean-field approximation.
(c) Log-log plots of the cumulative distributions of 
the relative population $y_i$ for $q=0.1$ and $n=10,000$.
We find that 
the tail of the distribution is lighter for stronger correlation. 
\label{f1}}
\end{figure}
Figure 1 shows the dependence of total growth rate $\log \bar{m}$ of 
$q$ and $r$.
The time average of $\log \bar{m}$ 
is calculated numerically in the case of $\tau_c=1$ (see Fig.~1(a)).
As we showed theoretically,  
the population has the same growth rate when $\phi=q(1-r)$ is identical.
There is a finite optimal value of $\phi=q(1-r)$, 
where the total growth rate has the maximal value (see Fig.~1(a)).
Figure 1(b) compares 
the time average of $\log \bar{m}$ obtained 
by the numerical simulation with the mean-field approximation
for some values of $\tau_c$ \eqref{e22}.
The mean-field approximation indicates 
that the total growth rate is a decreasing function 
of $\phi=q(1-r)$ for $\tau_c>0$.
If $\phi$ is not near 0, the numerical result agrees with 
the mean-field approximation very well.
On the other hand, for small $\phi$, 
the numerical average for $\log \bar{m}$ is smaller than 
the prediction.
The reason for this deviation is as follows.
If $\phi$ (also $q$) is near 0,
the distribution of the relative population size $y_i(t)$ has 
a heavy tail and the fluctuation of $\bar{m}$ is not negligible.
When the fluctuation of $\bar{m}(t)$ increases,
the time average of $\log \bar{m}(t)$ grows more smaller 
than the logarithm of the mean value of $\bar{m}(t)$ \cite{morita}.
As a result, if the environment has time correlation ($\tau_c>0$), 
there is a finite optimal value of $\phi$.
When the correlation time becomes longer,
the distribution's tail of $y_i(t)$ is lighter (see Fig.~1(c))
and the deviation from the mean-field approximation becomes smaller.
Thus, the optimal value of $\phi$ decreases as $\tau_c$ increases,
as seen in Fig.~1(b).

The result in Fig.~1(c) appears to be
in contrast to the previous works on 
Langevin systems with colored multiplicative noise,
where the distribution's tail is heavier due to temporal correlation
\cite{nakao,sato}. 
We focus on the simultaneous distribution of 
metapopulation coupled by migration, while
the previous works dealt one-dimension stochastic process.  
In our case, the temporal correlation tends to level off
the populations over all sites.

\begin{figure}
\begin{center}
\psfrag{q(1-r)}{{\small $\phi$}}
\psfrag{q}{{\small $q$}}
\psfrag{r}{{\small $r$}}
\includegraphics[width=0.5\textwidth]{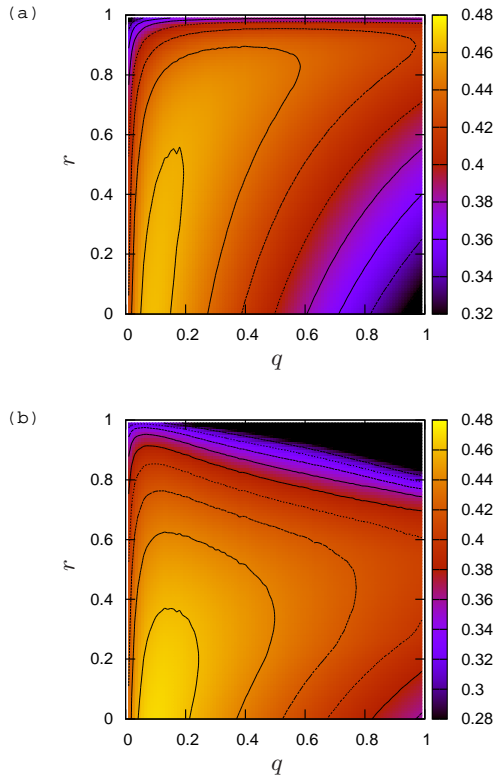}\\%
\end{center}
\caption{
The total growth rate $\log \bar{m}$ for 
two expanded cases as a function of $q$ and $r$.
(a) The case that 
transportation between sites has a cost.
The survival rate during the transportation is set to be $s=0.95$.
(b) The case that the preference is superlinear in the population,
where migrating individuals
select sites with the probability proportional to 
the square of the population size.
The black region means that the total growth rate
is smaller than the minimum value of the right color bar.
The other parameters are the same as in Fig.~1(b).
\label{f2}}
\end{figure}
Because of the degenerate character of the model,
the optimal values of $q$ and $r$ cannot be decided uniquely.
Here, we expand the model 
to resolve the degeneracy in two ways.
First, we take into account a cost of migration between sites.
Introduce the survival rate $s<1$, 
by which the second term of the right hand size of \eqref{e05}
is multiplied.
In this case, the total growth rate is
smaller than the one in the original model, when $q>0$. 
The difference between these two values increases with the migration rate $q$.
As a result, the values of $r$ for the optimal migration is zero.
Figure 2(a) shows the numerical plot of
the average of $\log \bar{m}$ for $s=0.95$.
In this case, the preference for fertile sites is not advantageous.

Second, we consider the case that 
the sensitivity of the preference is higher.
We assume that the migrating individuals follow 
\begin{equation} 
F(\{x_i(t)\})=r\frac{[m_i x_i(t)]^{\alpha}}{\sum_{j=1}^n[m_j x_j(t)]^{\alpha}}+(1-r)\frac{1}{n},
\end{equation} 
instead of \eqref{e04}.
Here, the exponent $\alpha$ represents the sensitivity of preference.
Figure 2(b) shows the numerical plot of
the average of $\log \bar{m}$ in the case that
the preference is superlinear ($\alpha>1$).
As seen in Fig.~2(b), the optimal value of $r$ is zero, again.
In the case of higher sensitivity,
because the population is excessively concentrated to particular sites, 
bet-hedging cannot work well. 
Conversely, if the preference is sublinear ($\alpha<1$),
the preference is favored.
Consequently, a high sensitivity makes the preference disadvantageous.

%
Up to here we have investigated a stochastic metapopulation model 
in fluctuating environments by comparing preferential dispersal
and random dispersals.
We showed that when the correlation time of the environments increases,
the optimal value of $\phi=q(1-r)$ decreases, i.e.,
the migration rate $q$ decreases for a given $r$, 
and the strength of the preference $r$ increase for a given $q$.
Moreover, if the migration has a cost or
the sensitivity of preference is high,
then random migration is favored compared with preferential migration 
These results are robust to variations the environmental
parameters $m_+$, $m_-$, $p$.
If we consider other shrewd preferences,
we may find the case that the preferential dispersal is advantageous.

There are similar studies 
that genetically clonal populations in fluctuating 
environment \cite{jansen,kussell}.
They contrasted two types of phenotype switchings: 
responsive switching and spontaneous random switching \cite{jansen,kussell}.
These studies showed that random switching is favored
rather than responsive switching in many cases.
The principle derived here may be the same with the above phenotype switchings.
We would like to emphasize that 
the principle we obtained here may
provide a reference for future research.
Since the bet hedging systems 
have been studied widely in ecology, evolution and economics,
we believe that our approach will be widely useful.

%
%
%

%



\begin{acknowledgments}
This work was supported by grants-in-aid from the Ministry of Education, 
Culture, Sports, Science and Technology of Japan to S.M. (No. 24500273) 
and J. Y. (No. 22370010 and No. 22255004). A part of the numerical 
computation in this work was carried out at the Yukawa Institute 
Computer Facility.
\end{acknowledgments}

\bibliography{ref.bib}

\begin{thebibliography}{16}%
\makeatletter
\providecommand \@ifxundefined [1]{%
 \@ifx{#1\undefined}
}%
\providecommand \@ifnum [1]{%
 \ifnum #1\expandafter \@firstoftwo
 \else \expandafter \@secondoftwo
 \fi
}%
\providecommand \@ifx [1]{%
 \ifx #1\expandafter \@firstoftwo
 \else \expandafter \@secondoftwo
 \fi
}%
\providecommand \natexlab [1]{#1}%
\providecommand \enquote  [1]{``#1''}%
\providecommand \bibnamefont  [1]{#1}%
\providecommand \bibfnamefont [1]{#1}%
\providecommand \citenamefont [1]{#1}%
\providecommand \href@noop [0]{\@secondoftwo}%
\providecommand \href [0]{\begingroup \@sanitize@url \@href}%
\providecommand \@href[1]{\@@startlink{#1}\@@href}%
\providecommand \@@href[1]{\endgroup#1\@@endlink}%
\providecommand \@sanitize@url [0]{\catcode `\\12\catcode `\$12\catcode
  `\&12\catcode `\#12\catcode `\^12\catcode `\_12\catcode `\%12\relax}%
\providecommand \@@startlink[1]{}%
\providecommand \@@endlink[0]{}%
\providecommand \url  [0]{\begingroup\@sanitize@url \@url }%
\providecommand \@url [1]{\endgroup\@href {#1}{\urlprefix }}%
\providecommand \urlprefix  [0]{URL }%
\providecommand \Eprint [0]{\href }%
\providecommand \doibase [0]{http://dx.doi.org/}%
\providecommand \selectlanguage [0]{\@gobble}%
\providecommand \bibinfo  [0]{\@secondoftwo}%
\providecommand \bibfield  [0]{\@secondoftwo}%
\providecommand \translation [1]{[#1]}%
\providecommand \BibitemOpen [0]{}%
\providecommand \bibitemStop [0]{}%
\providecommand \bibitemNoStop [0]{.\EOS\space}%
\providecommand \EOS [0]{\spacefactor3000\relax}%
\providecommand \BibitemShut  [1]{\csname bibitem#1\endcsname}%
\let\auto@bib@innerbib\@empty
\bibitem [{\citenamefont {Seger}\ and\ \citenamefont {Brockman}(1987)}]{seger}%
  \BibitemOpen
  \bibfield  {author} {\bibinfo {author} {\bibfnamefont {J.}~\bibnamefont
  {Seger}}\ and\ \bibinfo {author} {\bibfnamefont {H.~J.}\ \bibnamefont
  {Brockman}},\ }\enquote {\bibinfo {title} {What is bet-hedging?}}\ in\
  \href@noop {} {\emph {\bibinfo {booktitle} {Oxford Surveys in Evolutionary
  Biology}}},\ \bibinfo {editor} {edited by\ \bibinfo {editor} {\bibfnamefont
  {P.~H.}\ \bibnamefont {Harvey}}\ and\ \bibinfo {editor} {\bibfnamefont
  {L.}~\bibnamefont {Partridge}}}\ (\bibinfo  {publisher} {Oxford Univ.
  Press},\ \bibinfo {year} {1987})\ p.\ \bibinfo {pages} {182}\BibitemShut
  {NoStop}%
\bibitem [{\citenamefont {Levin}(1976)}]{r1}%
  \BibitemOpen
  \bibfield  {author} {\bibinfo {author} {\bibfnamefont {S.}~\bibnamefont
  {Levin}},\ }\href@noop {} {\bibfield  {journal} {\bibinfo  {journal} {Annu.\
  Rev.\ Ecol.\ Evol.\ Syst.}\ }\textbf {\bibinfo {volume} {7}},\ \bibinfo
  {pages} {287} (\bibinfo {year} {1976})}\BibitemShut {NoStop}%
\bibitem [{\citenamefont {Hastings}(1983)}]{r2}%
  \BibitemOpen
  \bibfield  {author} {\bibinfo {author} {\bibfnamefont {A.}~\bibnamefont
  {Hastings}},\ }\href@noop {} {\bibfield  {journal} {\bibinfo  {journal}
  {Theor.\ Popul.\ Biol.}\ }\textbf {\bibinfo {volume} {24}},\ \bibinfo {pages}
  {244} (\bibinfo {year} {1983})}\BibitemShut {NoStop}%
\bibitem [{\citenamefont {Haccou}\ and\ \citenamefont {Iwasa}(1995)}]{iwasa}%
  \BibitemOpen
  \bibfield  {author} {\bibinfo {author} {\bibfnamefont {P.}~\bibnamefont
  {Haccou}}\ and\ \bibinfo {author} {\bibfnamefont {Y.}~\bibnamefont {Iwasa}},\
  }\href@noop {} {\bibfield  {journal} {\bibinfo  {journal} {Theor. Popul.
  Biol.}\ }\textbf {\bibinfo {volume} {47}},\ \bibinfo {pages} {212} (\bibinfo
  {year} {1995})}\BibitemShut {NoStop}%
\bibitem [{\citenamefont {Ellner}\ and\ \citenamefont {Sasaki}(1995)}]{sasaki}%
  \BibitemOpen
  \bibfield  {author} {\bibinfo {author} {\bibfnamefont {S.}~\bibnamefont
  {Ellner}}\ and\ \bibinfo {author} {\bibfnamefont {A.}~\bibnamefont
  {Sasaki}},\ }\href@noop {} {\bibfield  {journal} {\bibinfo  {journal}
  {Evolution}\ }\textbf {\bibinfo {volume} {49}},\ \bibinfo {pages} {337}
  (\bibinfo {year} {1995})}\BibitemShut {NoStop}%
\bibitem [{\citenamefont {Childs}\ \emph {et~al.}(2010)\citenamefont {Childs},
  \citenamefont {Metcalf},\ and\ \citenamefont {Rees}}]{childs}%
  \BibitemOpen
  \bibfield  {author} {\bibinfo {author} {\bibfnamefont {D.~Z.}\ \bibnamefont
  {Childs}}, \bibinfo {author} {\bibfnamefont {C.~J.~E.}\ \bibnamefont
  {Metcalf}}, \ and\ \bibinfo {author} {\bibfnamefont {M.}~\bibnamefont
  {Rees}},\ }\href@noop {} {\  (\bibinfo {year} {2010})}\BibitemShut {NoStop}%
\bibitem [{\citenamefont {Williams}\ and\ \citenamefont {Hastings}(2011)}]{r3}%
  \BibitemOpen
  \bibfield  {author} {\bibinfo {author} {\bibfnamefont {P.~D.}\ \bibnamefont
  {Williams}}\ and\ \bibinfo {author} {\bibfnamefont {A.}~\bibnamefont
  {Hastings}},\ }\href@noop {} {\bibfield  {journal} {\bibinfo  {journal}
  {Proc.\ R.\ Soc.\ B}\ }\textbf {\bibinfo {volume} {278}},\ \bibinfo {pages}
  {1281} (\bibinfo {year} {2011})}\BibitemShut {NoStop}%
\bibitem [{\citenamefont {Jansen}\ and\ \citenamefont {Yoshimura}(1998)}]{r4}%
  \BibitemOpen
  \bibfield  {author} {\bibinfo {author} {\bibfnamefont {V.~A.~A.}\
  \bibnamefont {Jansen}}\ and\ \bibinfo {author} {\bibfnamefont
  {J.}~\bibnamefont {Yoshimura}},\ }\href@noop {} {\bibfield  {journal}
  {\bibinfo  {journal} {Proc.\ Natl.\ Acad.\ Sci.\ USA}\ }\textbf {\bibinfo
  {volume} {95}},\ \bibinfo {pages} {3696} (\bibinfo {year}
  {1998})}\BibitemShut {NoStop}%
\bibitem [{\citenamefont {Morita}\ and\ \citenamefont {Yoshimura}(2012)}]{r8}%
  \BibitemOpen
  \bibfield  {author} {\bibinfo {author} {\bibfnamefont {S.}~\bibnamefont
  {Morita}}\ and\ \bibinfo {author} {\bibfnamefont {J.}~\bibnamefont
  {Yoshimura}},\ }\href@noop {} {\bibfield  {journal} {\bibinfo  {journal}
  {Phys.\ Rev.\ E}\ }\textbf {\bibinfo {volume} {86}},\ \bibinfo {pages}
  {045102R} (\bibinfo {year} {2012})}\BibitemShut {NoStop}%
\bibitem [{\citenamefont {Luenberger}(1998)}]{r10}%
  \BibitemOpen
  \bibfield  {author} {\bibinfo {author} {\bibfnamefont {D.~G.}\ \bibnamefont
  {Luenberger}},\ }\href@noop {} {\emph {\bibinfo {title} {Investment
  Science}}}\ (\bibinfo  {publisher} {Prentice Hall, Englewood Cliffs, NJ},\
  \bibinfo {year} {1998})\BibitemShut {NoStop}%
\bibitem [{\citenamefont {Schreiber}(2010)}]{r7}%
  \BibitemOpen
  \bibfield  {author} {\bibinfo {author} {\bibfnamefont {S.~J.}\ \bibnamefont
  {Schreiber}},\ }\href@noop {} {\bibfield  {journal} {\bibinfo  {journal}
  {Proc.\ R.\ Soc.\ London, Ser.\ B}\ }\textbf {\bibinfo {volume} {277}},\
  \bibinfo {pages} {1907} (\bibinfo {year} {2010})}\BibitemShut {NoStop}%
\bibitem [{\citenamefont {Morita}\ and\ \citenamefont
  {Yoshimura}(2013)}]{morita}%
  \BibitemOpen
  \bibfield  {author} {\bibinfo {author} {\bibfnamefont {S.}~\bibnamefont
  {Morita}}\ and\ \bibinfo {author} {\bibfnamefont {J.}~\bibnamefont
  {Yoshimura}},\ }\href@noop {} {\bibfield  {journal} {\bibinfo  {journal}
  {Phys.\ Rev.\ E}\ }\textbf {\bibinfo {volume} {88}},\ \bibinfo {pages}
  {052809} (\bibinfo {year} {2013})}\BibitemShut {NoStop}%
\bibitem [{\citenamefont {Nakao}(1998)}]{nakao}%
  \BibitemOpen
  \bibfield  {author} {\bibinfo {author} {\bibfnamefont {H.}~\bibnamefont
  {Nakao}},\ }\href@noop {} {\bibfield  {journal} {\bibinfo  {journal} {Phys.\
  Rev.\ E}\ }\textbf {\bibinfo {volume} {58}},\ \bibinfo {pages} {1591}
  (\bibinfo {year} {1998})}\BibitemShut {NoStop}%
\bibitem [{\citenamefont {Sato}\ \emph {et~al.}(2000)\citenamefont {Sato},
  \citenamefont {Takayasu},\ and\ \citenamefont {Sawada}}]{sato}%
  \BibitemOpen
  \bibfield  {author} {\bibinfo {author} {\bibfnamefont {A.-H.}\ \bibnamefont
  {Sato}}, \bibinfo {author} {\bibfnamefont {H.}~\bibnamefont {Takayasu}}, \
  and\ \bibinfo {author} {\bibfnamefont {Y.}~\bibnamefont {Sawada}},\
  }\href@noop {} {\bibfield  {journal} {\bibinfo  {journal} {Phys.\ Rev.\ E}\
  }\textbf {\bibinfo {volume} {61}},\ \bibinfo {pages} {1081} (\bibinfo {year}
  {2000})}\BibitemShut {NoStop}%
\bibitem [{\citenamefont {Jansen}\ and\ \citenamefont {Stumpf}(2005)}]{jansen}%
  \BibitemOpen
  \bibfield  {author} {\bibinfo {author} {\bibfnamefont {V.~A.~A.}\
  \bibnamefont {Jansen}}\ and\ \bibinfo {author} {\bibfnamefont {M.~P.~H.}\
  \bibnamefont {Stumpf}},\ }\href@noop {} {\bibfield  {journal} {\bibinfo
  {journal} {Science}\ }\textbf {\bibinfo {volume} {309}},\ \bibinfo {pages}
  {2005} (\bibinfo {year} {2005})}\BibitemShut {NoStop}%
\bibitem [{\citenamefont {Kussell}\ and\ \citenamefont
  {Leibler}(2005)}]{kussell}%
  \BibitemOpen
  \bibfield  {author} {\bibinfo {author} {\bibfnamefont {E.}~\bibnamefont
  {Kussell}}\ and\ \bibinfo {author} {\bibfnamefont {S.}~\bibnamefont
  {Leibler}},\ }\href@noop {} {\bibfield  {journal} {\bibinfo  {journal}
  {Science}\ }\textbf {\bibinfo {volume} {309}},\ \bibinfo {pages} {2075}
  (\bibinfo {year} {2005})}\BibitemShut {NoStop}%
\end{thebibliography}%

\end{document}